\documentclass[11pt]{article}
\usepackage{hyperref}
\usepackage{natbib}
\usepackage{amsmath}
\usepackage{nicefrac}
\usepackage[usenames,dvipsnames]{xcolor}
\usepackage{graphicx}
\usepackage{footnote}
\usepackage{rotating}
\usepackage{slashbox}
\usepackage{afterpage}
\usepackage{float}
\newcommand\pubnumber{Article 14 in eConf C1304143}
\newcommand\pubdate{\today}

\def\affiliation{Department of Physics \& Institute for Fusion Studies\\
The University of Texas at Austin, TX 78712, USA; amir@physics.utexas.edu}

\def\gtrsim{\mathrel{\hbox{\rlap{\hbox{\lower4pt\hbox{$\sim$}}}\hbox{$>$}}}}
\def\lessim{\mathrel{\hbox{\rlap{\hbox{\lower4pt\hbox{$\sim$}}}\hbox{$<$}}}}
\newcommand{\liso}{L_{iso}}
\newcommand{\eiso}{E_{iso}}
\newcommand{\epkz}{E_{p,z}}
\newcommand{\durz}{T_{90,z}}
\newcommand{\pbol}{P_{bol}}
\newcommand{\sbol}{S_{bol}}
\newcommand{\epk}{E_{p}}
\newcommand{\dur}{T_{90}}

\def\Title#1{\begin{center} {\Large #1 } \end{center}}
\def\Author#1{\begin{center}{ \sc #1} \end{center}}
\def\Address#1{\begin{center}{ \it #1} \end{center}}

\newcommand\pubblock{\rightline{\begin{tabular}{l} \pubnumber\\
         \pubdate  \end{tabular}}}
\newenvironment{Abstract}{\begin{quotation}  }{\end{quotation}}
\newenvironment{Presented}{\begin{quotation} \begin{center}
             PRESENTED AT\end{center}\bigskip
      \begin{center}\begin{large}}{\end{large}\end{center} \end{quotation}}
\def\Acknowledgements{\bigskip  \bigskip \begin{center} \begin{large}
             \bf ACKNOWLEDGEMENTS \end{large}\end{center}}
\begin{document}
\begin{titlepage}
\pubblock

\vfill
\Title{Gamma-Ray bursts: Energetics and Prompt Correlations}
\vfill
\Author{Amir Shahmoradi}
\Address{\affiliation}
\vfill
\begin{Abstract}
    A model is presented here that is capable of simultaneously describing the luminosity function and the underlying joint population distribution of the prompt spectral and temporal parameters of Gamma-Ray Bursts (GRBs), subject to the detection threshold of gamma-ray instruments--in particular, BATSE and Fermi. Based on $2130$ GRB prompt emission data in the BATSE catalog, I show that the population properties of the two classes of GRBs -- Long \& Short durations -- bear striking similarities in the 4-dimensional space of prompt parameters: peak luminosity ($\liso$), total isotropic emission ($\eiso$), time-integrated spectral peak energy ($\epkz$) \& the prompt duration ($\durz$). The two well-known Amati ($\eiso-\epkz$) \& Yonetoku ($\liso-\epkz$) relations are shown to be highly affected by selection effects, undermining the legitimacy of their frequent uses in the studies of Dark Energy's equation of state and different cosmological models. In particular, I show that the slope of the Amati relation is likely in the range $0.2-0.3$, corresponding to a Pearson's correlation strength of $\rho=0.58\pm0.04$. This predicted slope is significantly ($>16\sigma$) less than the value currently perceived by the GRB community ($\sim0.56$). I argue that similar $\eiso-\epkz$ \& $\liso-\epkz$ relations with approximately same strength and significance should exist in the population of Short GRBs. Also predicted by the model is the strong {\it positive} correlation of the prompt emission duration (e.g., $\durz$) with $\eiso$ and $\liso$ in {\it both} classes of Short and Long GRBs.
\end{Abstract}
\vfill
\begin{Presented}
$7^{th}$ Huntsville Gamma-Ray Burst Symposium
\\
Nashville, TN,  April 14--18, 2013
\end{Presented}
\vfill
\end{titlepage}
\def\thefootnote{\fnsymbol{footnote}}
\setcounter{footnote}{0}

\section{Introduction}

\begin{figure}[t]
        \centering
        \begin{tabular}{cc}
            \includegraphics[scale=0.55]{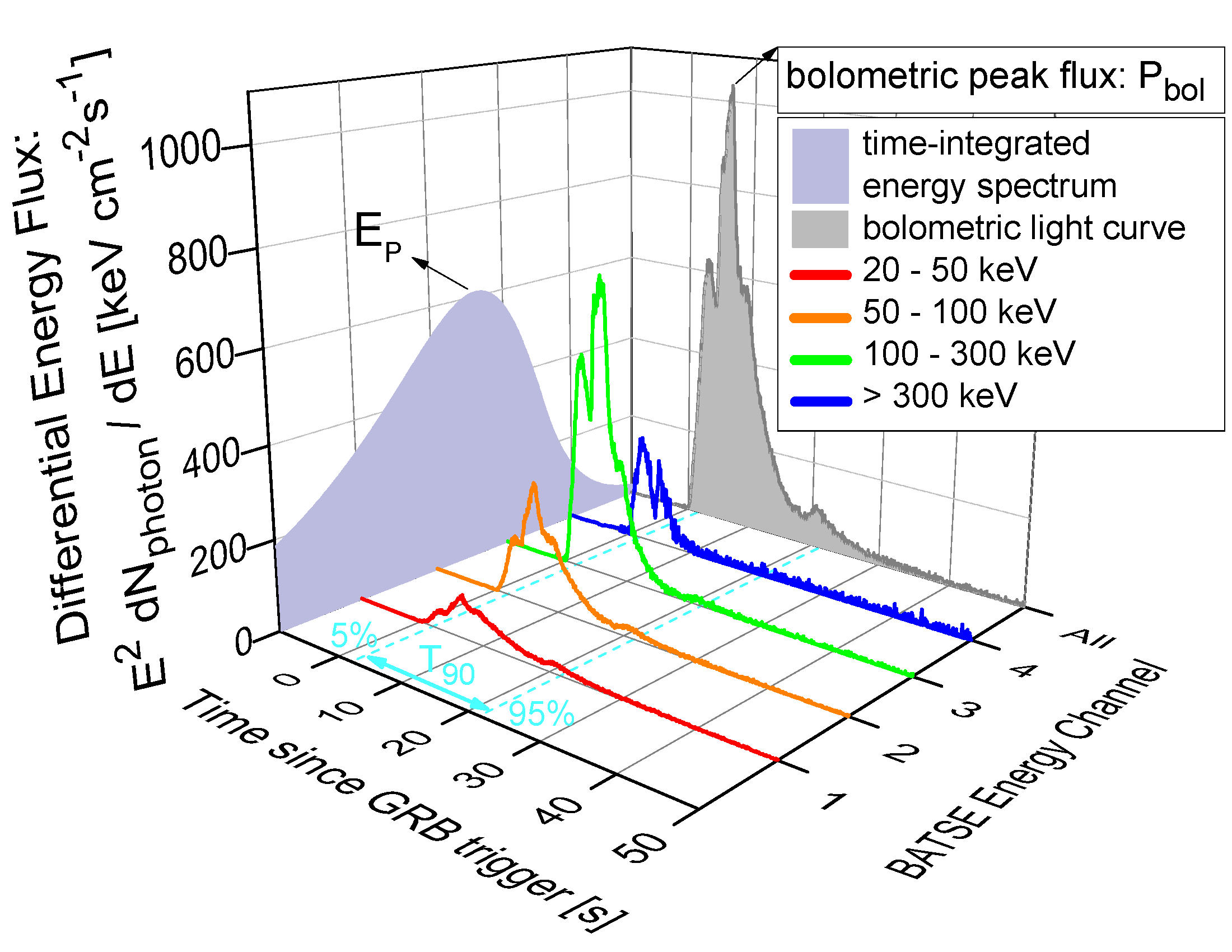}
        \end{tabular}
        \caption{\small An example lightcurve of GRB prompt $\gamma$-ray emission (BATSE GRB trigger 1085) in different energy bands, corresponding to BATSE 4 main energy channels. The graph depicts the definitions of four main observer-frame GRB prompt variables that are used in the population study presented in this work: The bolometric peak flux ($\pbol$), fluence ($\sbol$), the spectral peak energy ($\epk$) \& the prompt duration (i.e., $\dur$). It will be later argued in this work that {\bf after a careful consideration of detector threshold in the analysis, the joint population distribution of these four prompt variables in GRB's rest-frame closely resembles a 4-dimensional multivariate log-normal distribution for both classes of Short and Long GRBs}. \label{fig:3dgraph}}
    \end{figure}

    Despite significant progress over the past decade, difficulties in modeling the complex effects of detector threshold on the multivariate population distribution of the prompt-emission properties and the lack of a sufficiently large sample of uniformly detected GRBs has led the community to focus on individual prompt spectral$/$temporal variables, most importantly, on the Luminosity Function  ({\bf LF}) of GRBs.  A more accurate modeling of the LF, however, requires {\it at least} two variables incorporated in the LF of Long GRBs ({\bf LGRBs}): the bolometric peak flux ($\pbol$) and the observed peak energy ($\epk$). The parameter $\epk$ is required, since most $\gamma$-ray detectors are photon counters, a quantity that depends on not only $\pbol$, but also $\epk$ of the burst. This leads to the requirement of using a {\bf bivariate} distribution as the minimum acceptable statistical model to begin with, for the purpose of constraining LGRBs' luminosity function. For Short class of GRBs ({\bf SGRBs}), the joint {\bf trivariate} distribution of $\pbol$, $\epk$ and the observed duration (e.g., $\dur$) is the minimum acceptable model in order to correctly account for the detection threshold of most $\gamma$-ray detectors. The goal of the presented analysis is,

    \begin{itemize}
        \item
            to provide a quantitative phenomenological classification method for GRBs based on the observed prompt $\gamma$-ray emission properties, independent of detector specifications and limitations.
        \item
            to derive a multivariate statistical model that is capable of reproducing the luminosity function, energetics, duration distributions and the true underlying correlations among four main parameters of the prompt $\gamma$-ray emission in both classes of SGRBs \& LGRBs:  {\bf $\liso$, $\eiso$, $\epkz$ \& $\durz$} (Figure \ref{fig:3dgraph}) while paying careful attention to selections effects and observational biases that might affect GRB data.

        \item
            to gauge the utility and strength of the claimed high-energy correlations among the spectral parameters of GRBs in cosmological studies, in particular, the study of the Dark Energy's equation of state at high redshifts.
    \end{itemize}

    \begin{figure}[t]
        \centering
        \begin{tabular}{cc}
            \includegraphics[scale=0.3]{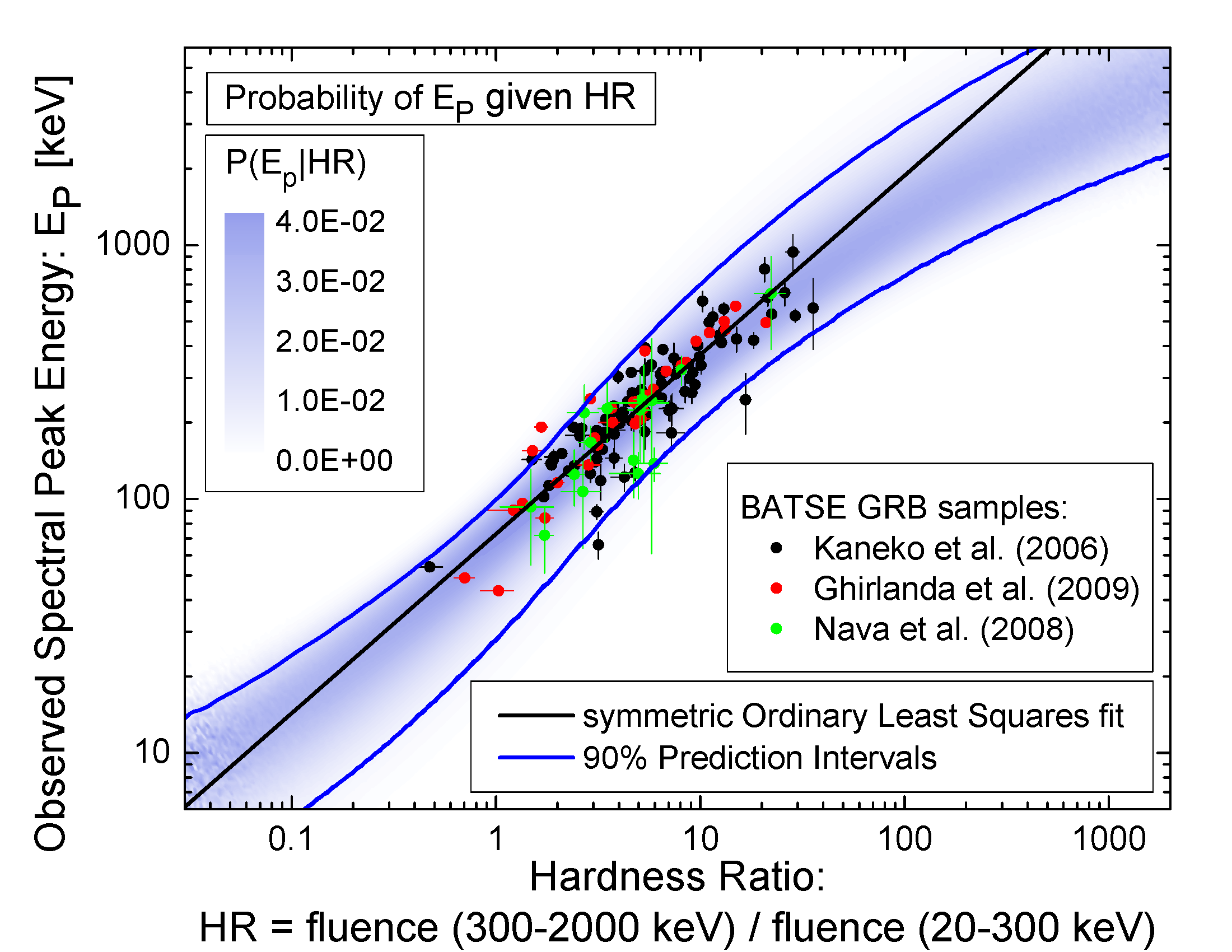} & \includegraphics[scale=0.3]{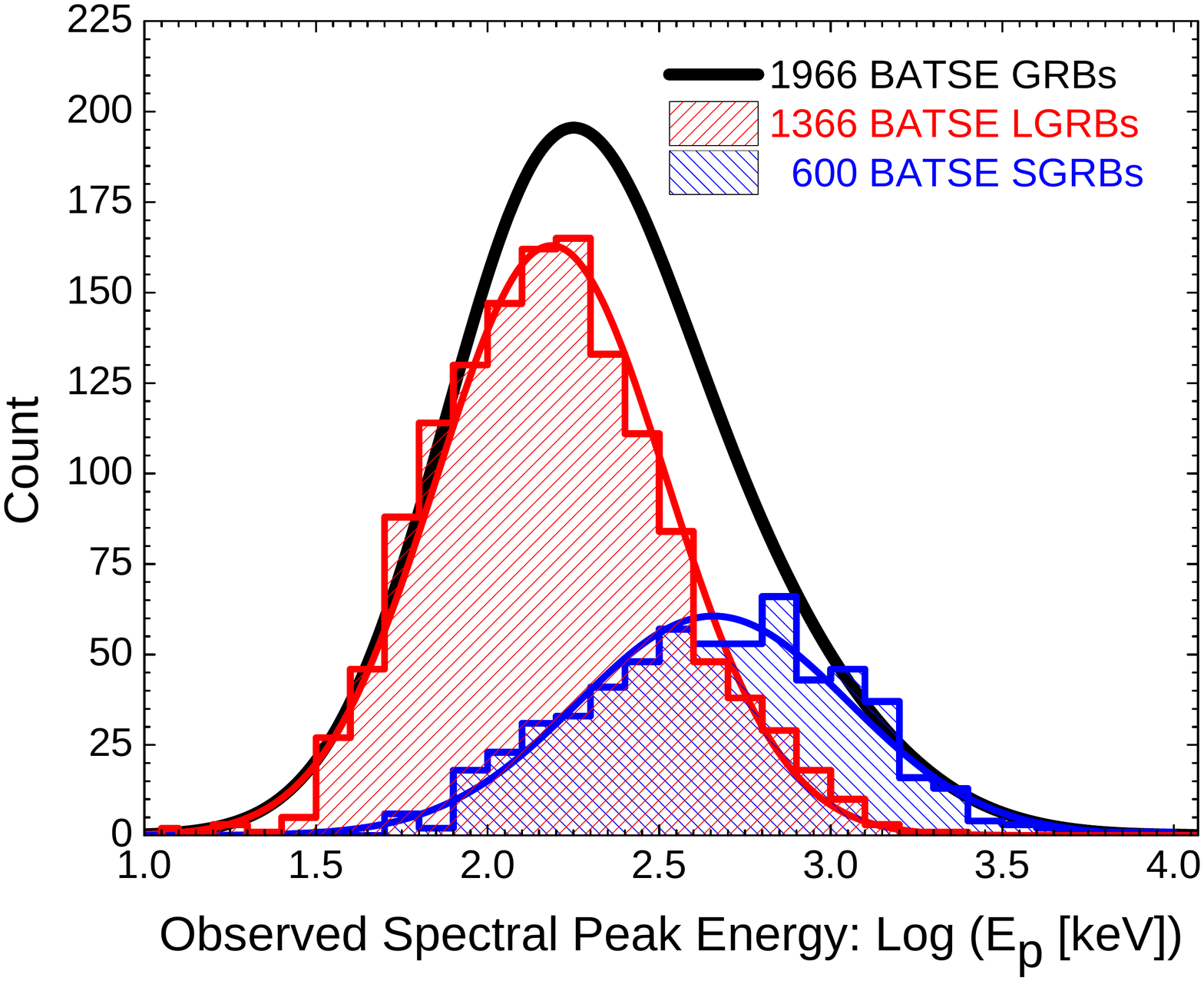} \\
            \includegraphics[scale=0.3]{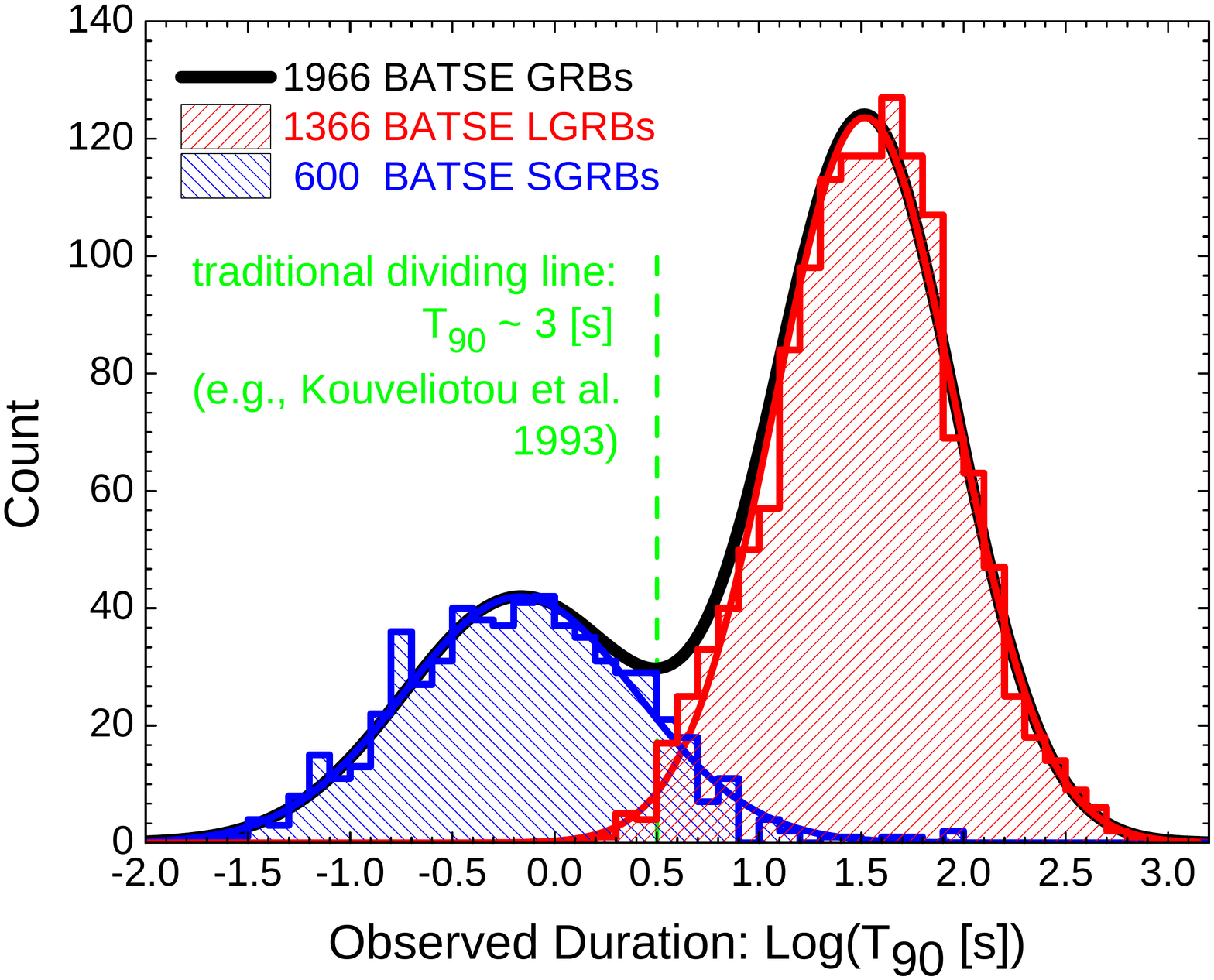} & \includegraphics[scale=0.3]{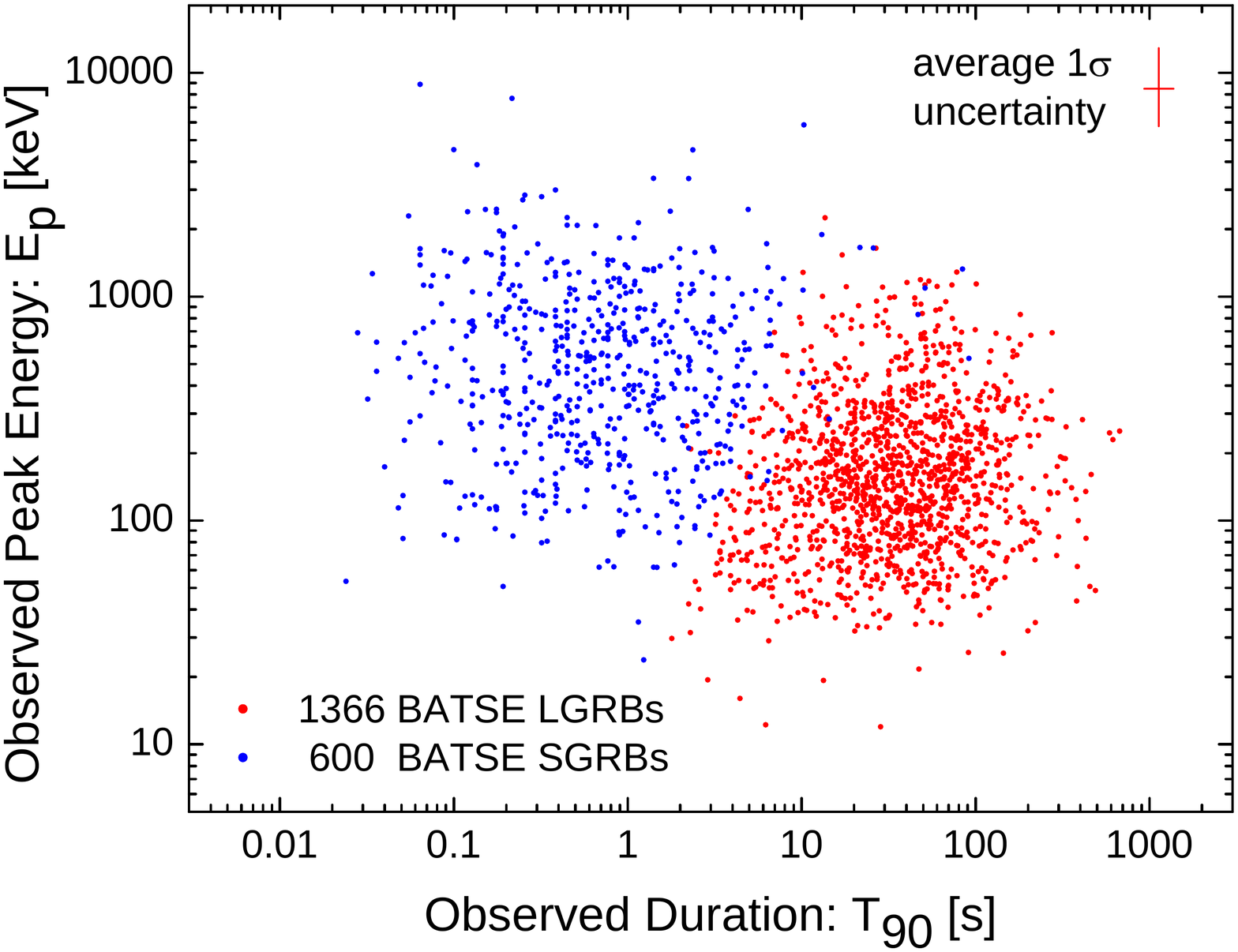}
        \end{tabular}
        \caption{\small{\bf Top Left}: Example graph depicting the strong correlation of $\epk$ with Hardness Ratio (HR) of BATSE GRBs \citep[c.f.,][for details]{shahmoradi_hardness_2010}. {\bf Top Right}: Derived $\epk$ distribution of 1966 BATSE GRBs used in the population study of this work, based on HR--$\epk$ relation of \citet{shahmoradi_hardness_2010}. {\bf Bottom Left}: $\dur$ distribution of 1966 BATSE GRBs. The $\epk$ estimates of 2130 BATSE catalog GRBs are publicly available for download at: {\footnotesize\url{https://sites.google.com/site/amshportal/research/aca/in-the-news/lgrb-world-model}} {\bf Bottom Right}: The joint $\dur-\epk$ distribution of 1966 BATSE GRBs, classified according to fuzzy C-means clustering algorithm \citep[c.f., Sec. \ref{sec:sampleselection}; also][Sec. 2.1 \& App. A therein for details]{shahmoradi_multivariate_2013}. \label{fig:classification}}
    \end{figure}

    Towards the above goals, this work focuses on the largest catalog of GRBs available to this date: the BATSE catalog of 2130 GRBs, though it can be readily expanded to other GRB databases, in particular the Fermi catalog of GRBs. Much of the methods and results presented in this paper have been already worked out in a series of articles recently published by the author \citep[][]{shahmoradi_how_2009, shahmoradi_hardness_2010, shahmoradi_possible_2011, shahmoradi_multivariate_2013} or is in preparation for publication (Shahmoradi \& Nemiroff 2013).\footnote{Data for BATSE GRBs with firmly measured peak flux, fluence and duration are taken from The BATSE Gamma Ray Burst Catalogs: \url{http://www.batse.msfc.nasa.gov/batse/grb/catalog/}. The spectral peak energy ($\epk$) estimates of these events are taken from \citet{shahmoradi_hardness_2010}, also available for download at: \url{https://sites.google.com/site/amshportal/research/aca/in-the-news/lgrb-world-model}.} 

\section{Sample Selection}
\label{sec:sampleselection}

    The traditional definition of GRB classes is based on a sharp cutoff in the observed duration ($\dur$) distribution of GRBs, generally set at $\dur\sim 2-3[s]$ \citep[e.g.,][]{kouveliotou_identification_1993}. Here, to ensure the least amount of bias in classification and a correct analysis, BATSE GRBs are classified according to fuzzy C-means clustering algorithm, based on two GRB observable: $\dur$ \& $\epk$ as shown in Figure \ref{fig:classification} below. Unlike $\dur$ \& $\epk$ which are weakly coupled to (i.e., correlated with) the variable $\pbol$, the population distribution of the two other GRB prompt variables ($\sbol$ \& $\pbol$) are strongly affected by the detector thresholds (c.f., Figure \ref{fig:bivariate}) and are not suitable for classifications based on fuzzy C-means algorithms. This is mainly due to the sensitivity of C-means clustering method to different subgroup sizes, orientations, and asymmetries \citep[c.f.,][Sec. 2.1 \& Appendix A therein for details]{shahmoradi_multivariate_2013}.

\section{Model Construction \& Fitting}
\label{sec:MCF}

    Here, the multivariate log-normal distribution is proposed as the simplest natural candidate model, capable of describing data. The motivation behind this choice of model comes from the available observational data that closely resemble a joint multivariate log-normal distribution for the four most widely studied temporal and spectral parameters of GRBs in the observer frame: $\pbol$, $\sbol$, $\epk$ \& $\dur$, truncated by BATSE Large Area Detector ({\bf LAD}) threshold:  since most GRBs originate from moderate redshifts $z\sim1-–3$, a fact known thanks to Swift satellite \citep[e.g.,][]{butler_cosmic_2010}, the convolution of these observer-frame parameters with the redshift distribution results in negligible variation in the shape of the rest-frame joint distribution of the same GRB parameters.

    \begin{figure}[t!]
        \centering
        \begin{tabular}{cc}
            \includegraphics[scale=0.3]{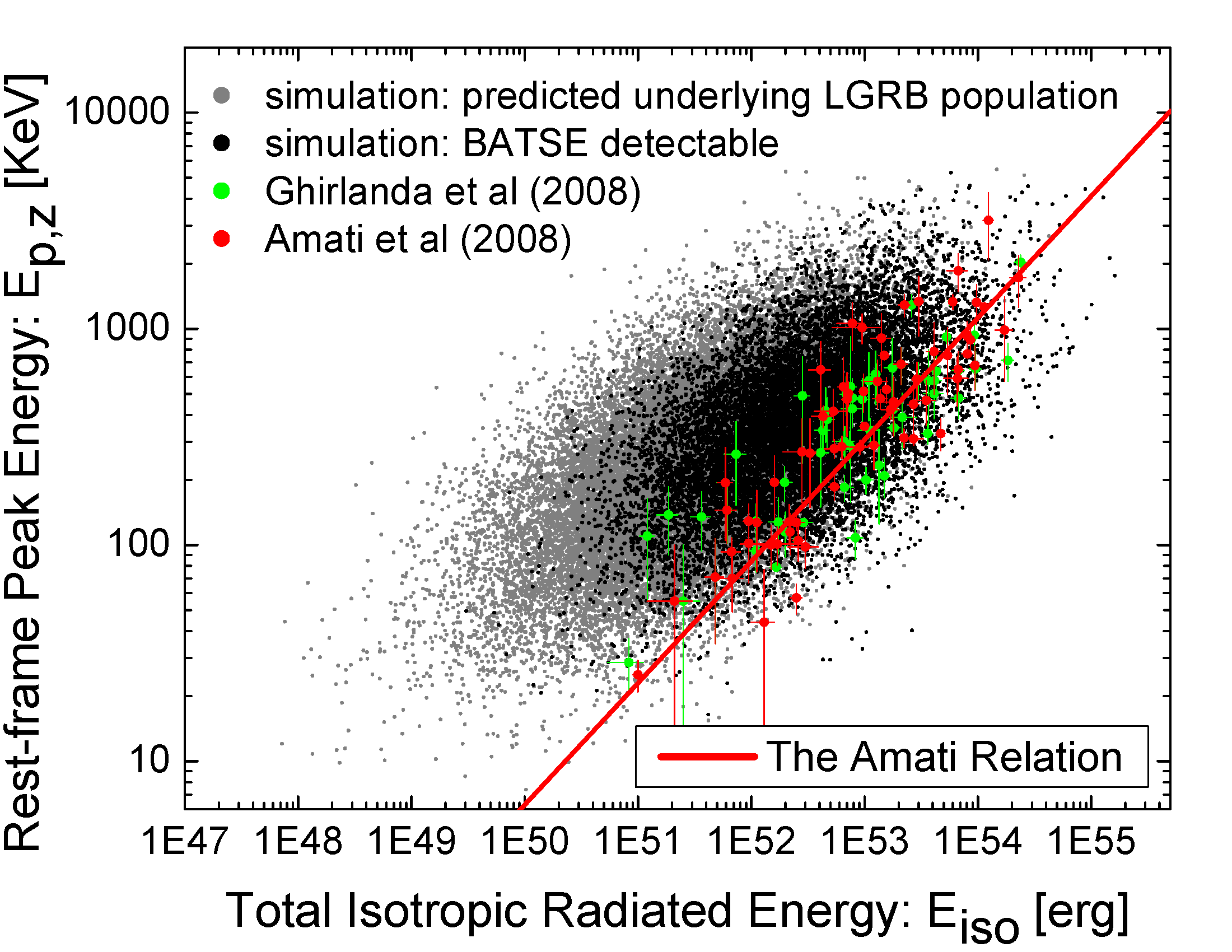} & \includegraphics[scale=0.3]{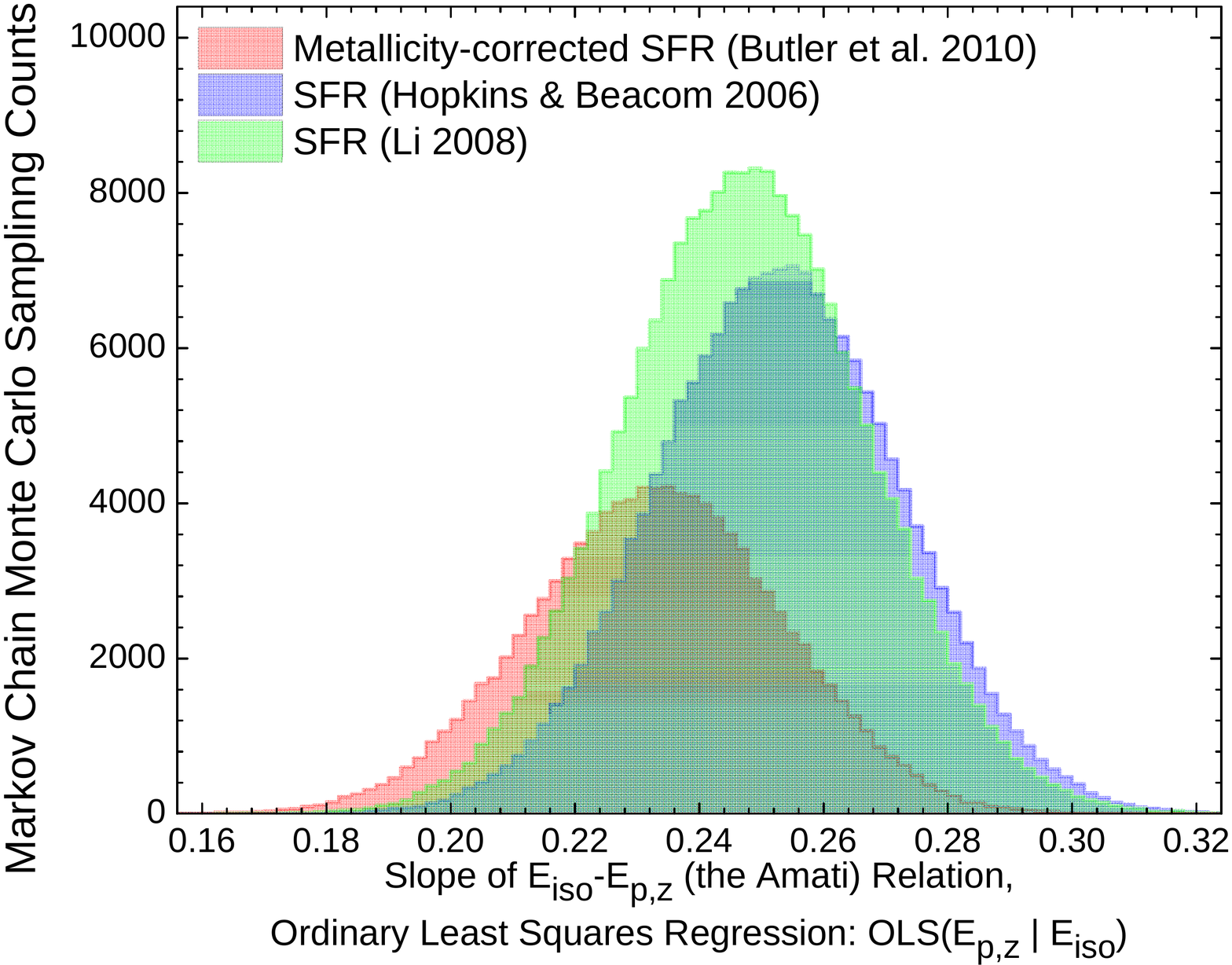} \\
        \end{tabular}
        \caption{\small{\bf Left}: Predictions of the LGRB world model of \citet{shahmoradi_multivariate_2013} for one of the most highly debated correlations in GRB literature: $\epk-\eiso$ (the Amati) relation. Complex selection effects on redshift measurement, spectral analysis and detection process result in a tight relation, while the predicted underlying population of LGRBs by the model show a significantly wider dispersion than what is currently realized by the GRB community. An $\epk-\eiso$ correlation of similar strength \& dispersion is also predicted for Short class of GRBs (Shahmoradi \& Nemiroff 2013, in preparation).  {\bf Right}: The predicted slope of the Amati relation according to three different redshift scenarios considered for BATSE LGRBs: Star Formations Rate (SFR) of \citet{hopkins_normalization_2006}, the SFR of \citet{li_star_2008}, LGRBs tracing the metallicity evolution as derived from Swift catalog by \citet{butler_cosmic_2010}. {\bf The predicted Ordinary Least Squares (OLS) regression slope of the Amati relation ($\sim0.2-0.3$) based on BATSE catalog of 1366 LGRBs is in stark contrast with the commonly accepted Amati relation \citep[e.g.,][]{amati_measuring_2008, ghirlanda_e_2008} with an OLS regression slope of $\sim0.56$}. \label{fig:amati}}
    \end{figure}

    The process of GRB observation is therefore considered as a non-homogeneous Poisson process whose mean rate parameter (i.e.,  the cosmic GRB differential rate), is the product of the differential comoving GRB rate density (the redshift distribution of GRBs) with a 4-dimensional log-normal probability density function with each dimension corresponding to one prompt variable in GRB's rest-frame:  $\liso$, $\eiso$, $\epkz$, $\durz$ \citep[c.f., Sec. 2.2 in][for details of LGRB world model; also Shahmoradi \& Nemiroff 2013, in preparation, for details of SGRBs world model]{shahmoradi_multivariate_2013}. The observed rate of GRBs, is then the result of the convolution of the cosmic GRB rate with BATSE LAD threshold. The probability of detection for an LGRB is modeled by the cumulative density function of log-normal distribution as described in Appendix B of \citet{shahmoradi_multivariate_2013}. Based on the observation that almost all 1366 BATSE LGRBs have durations of $\dur>1~[s]$ the primary trigger timescale for BATSE LGRBs is assumed to be $1024~[ms]$. This eliminates the relatively complex dependence of the detection probability on the duration of the LGRB events. For SGRBs, the situation is nontrivial and will be discussed by Shahmoradi \& Nemiroff (2013, in preparation).

    The best-fit parameters are obtained by the method of maximum likelihood. This is done by maximizing the likelihood function of the model, given the observational data, using a variant of the Metropolis–-Hastings Markov Chain Monte Carlo ({\bf MCMC}) algorithm. Details of constructing the objective (likelihood) function are given in Appendix C of \citet{shahmoradi_multivariate_2013} for LGRBs world model and will be provided in Shahmoradi \& Nemiroff (2013, in preparation) for SGRBs world model. Since almost no redshift information is available for BATSE catalog of GRBs, the probability for each LGRB in the likelihood function is marginalized over all possible redshifts in the range $z\in[0.1,\infty]$. To reduce the simulation runtime, all algorithms including MCMC are implemented in Fortran (2008 standard).

\section{Results \& Concluding Remarks}
\label{sec:RCR}

    An example of fitting results and the LGRB world model prediction for one of the most debated relations is presented in Figure \ref{fig:amati}. Also, the predicted underlying population distribution of Short and Long GRBs together with BATSE 1366 LGRBs \& 600 SGRBs data in the observer-frame are depicted in Figure \ref{fig:bivariate} \citep[c.f.,][Sec. 2.4 \& Figures 2--6 for extensive goodness-of-fit tests]{shahmoradi_multivariate_2013}.

    Summarized below are the principal conclusions drawn from the analysis based on the proposed world models for Short- \& Long-duration classes of GRBs.

    \afterpage{
    \begin{figure}[H]
        \centering
        \includegraphics[scale=0.0792]{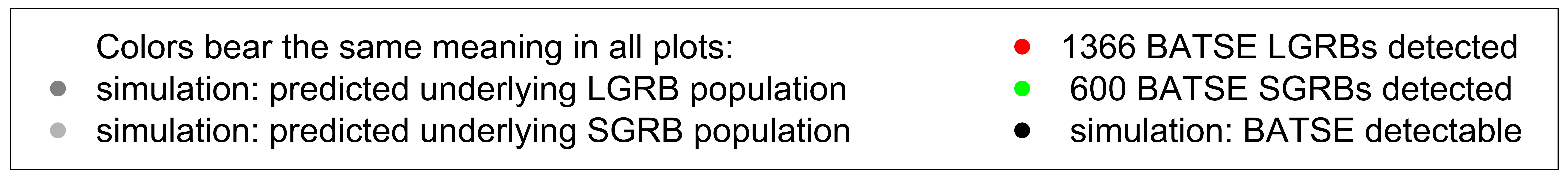}
        \begin{tabular}{cc}
            \includegraphics[scale=0.309]{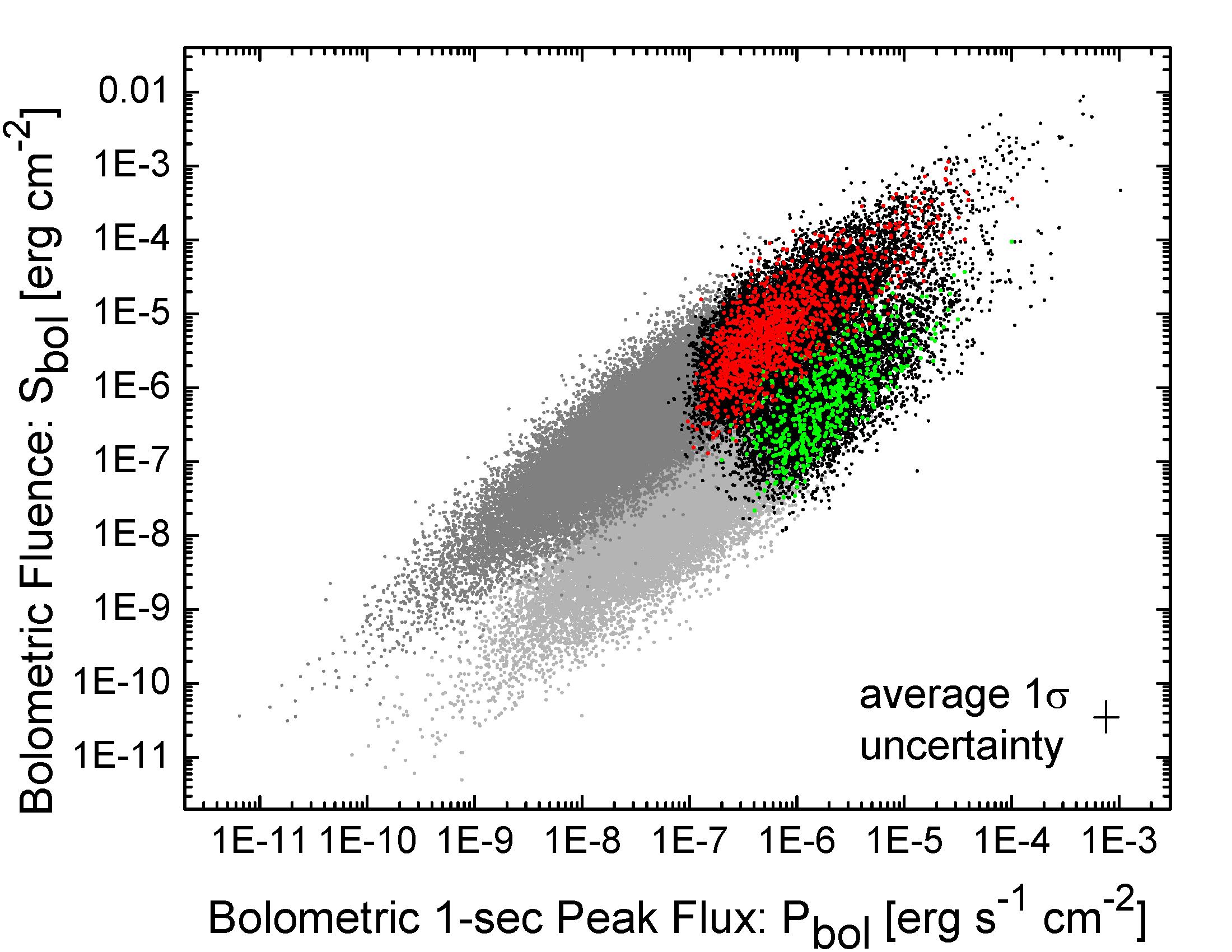} & \includegraphics[scale=0.309]{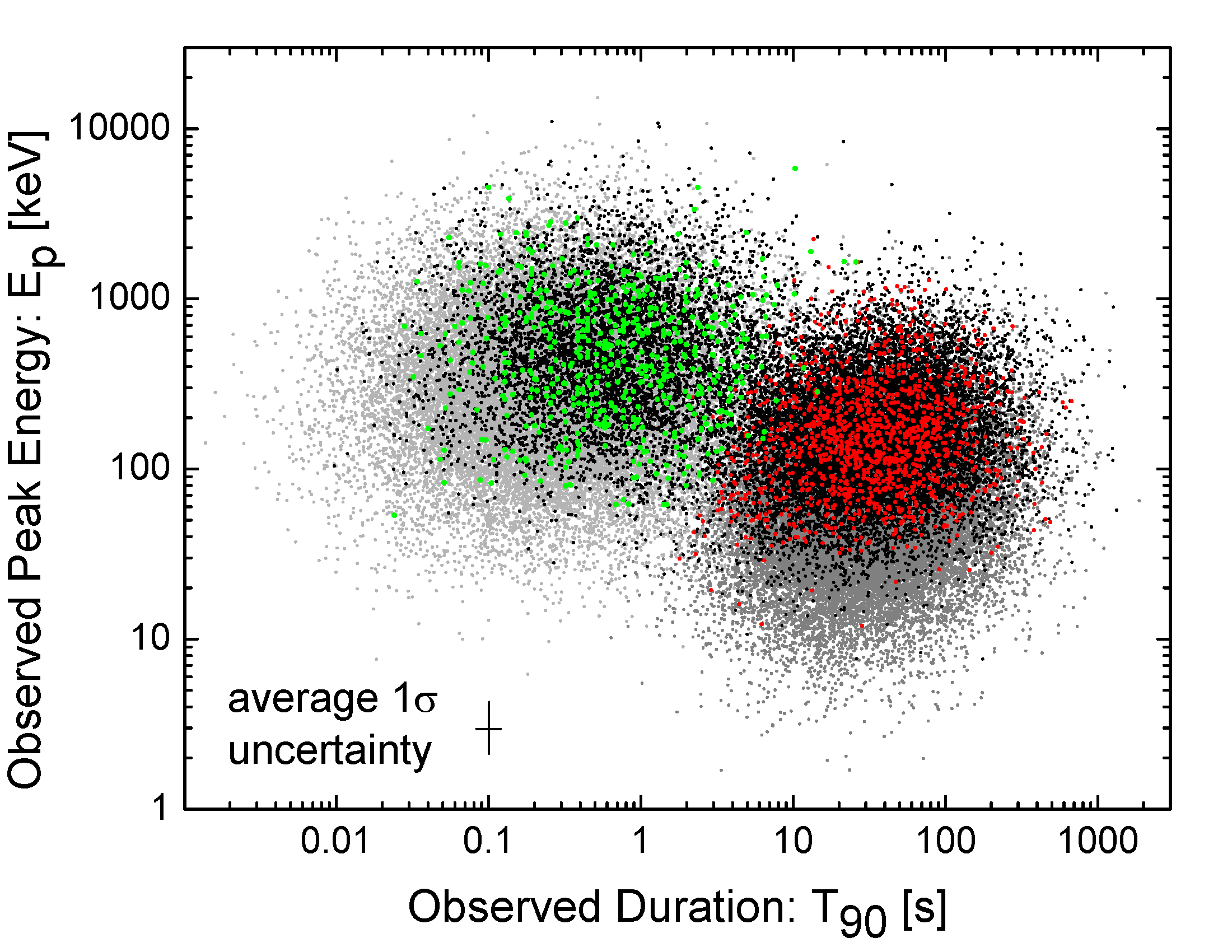} \\
            \includegraphics[scale=0.309]{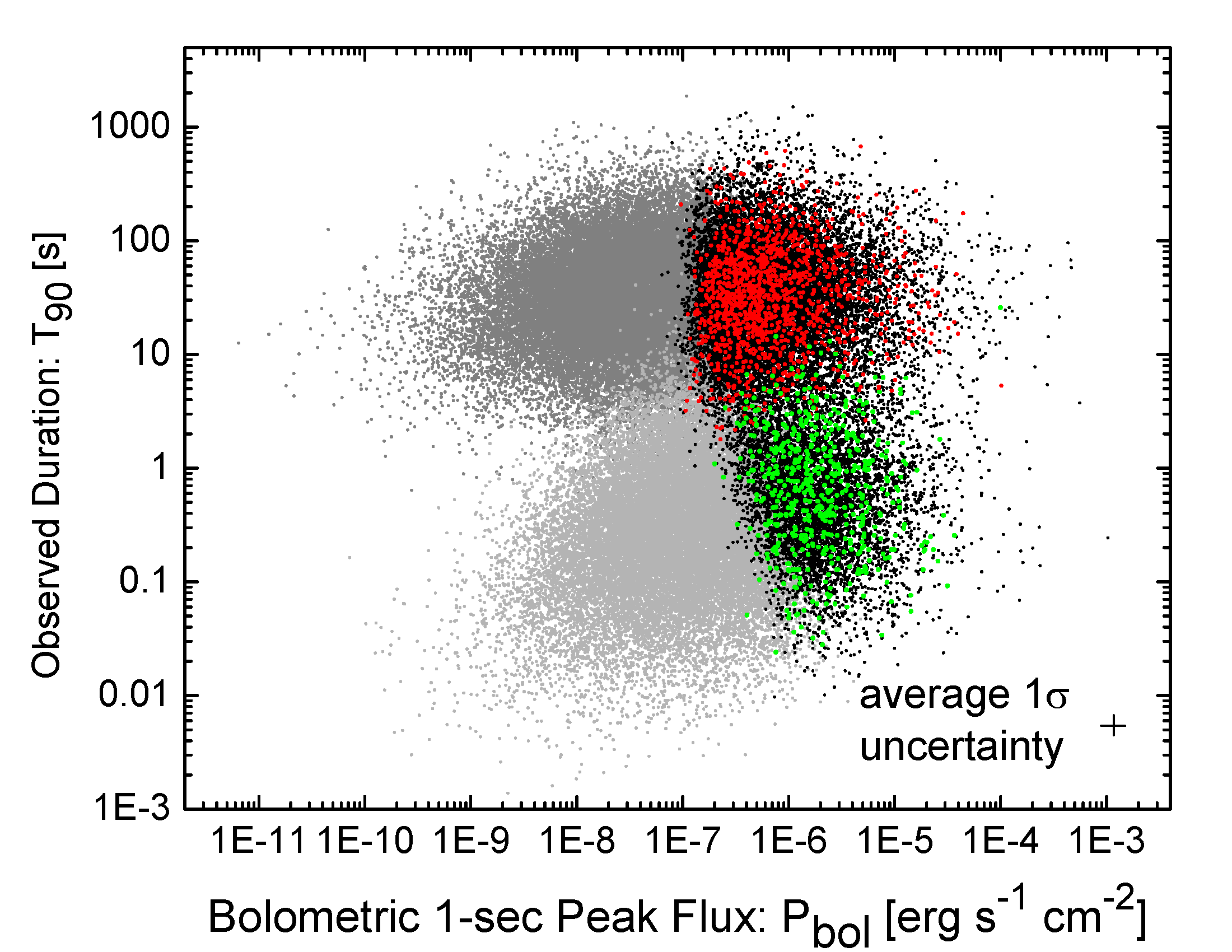} & \includegraphics[scale=0.309]{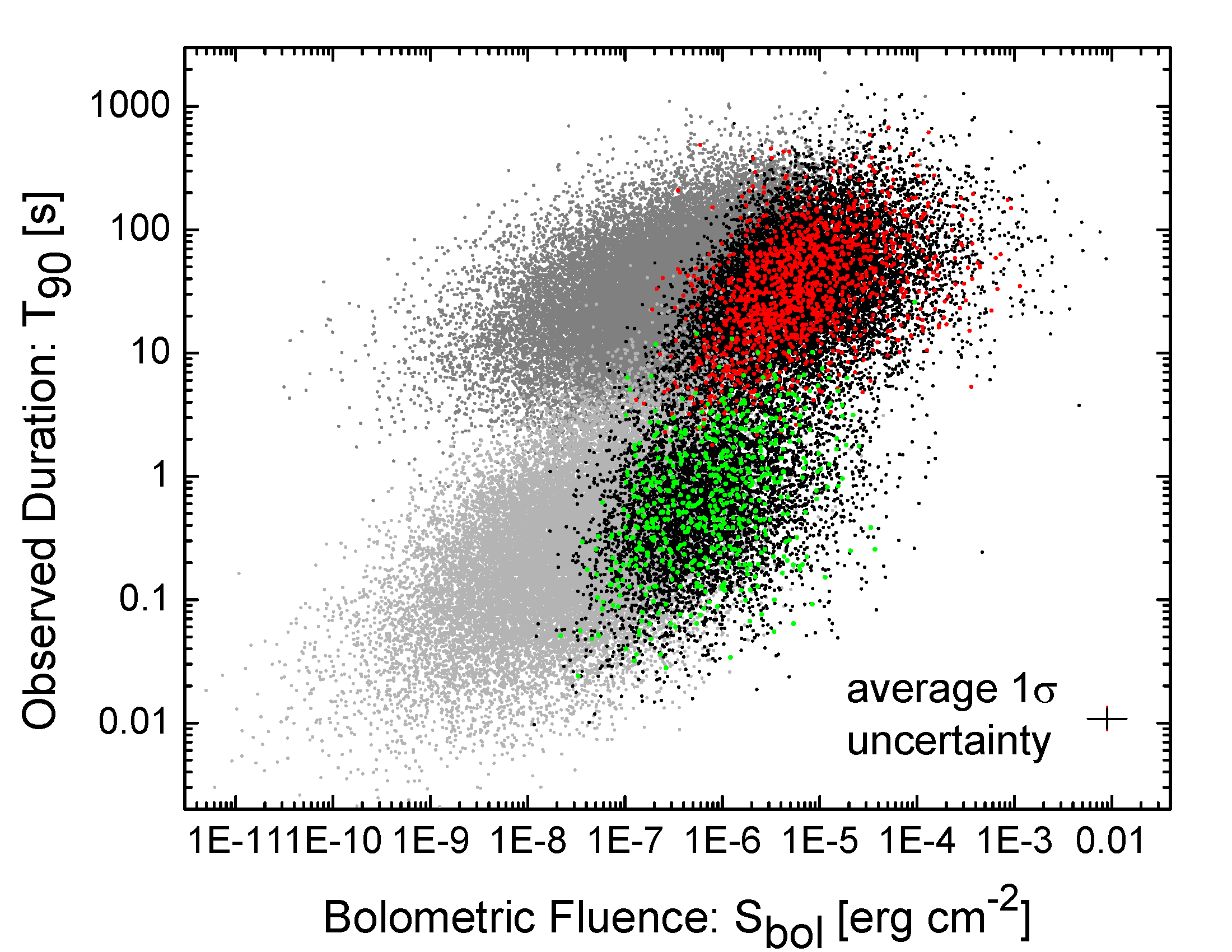} \\
            \includegraphics[scale=0.309]{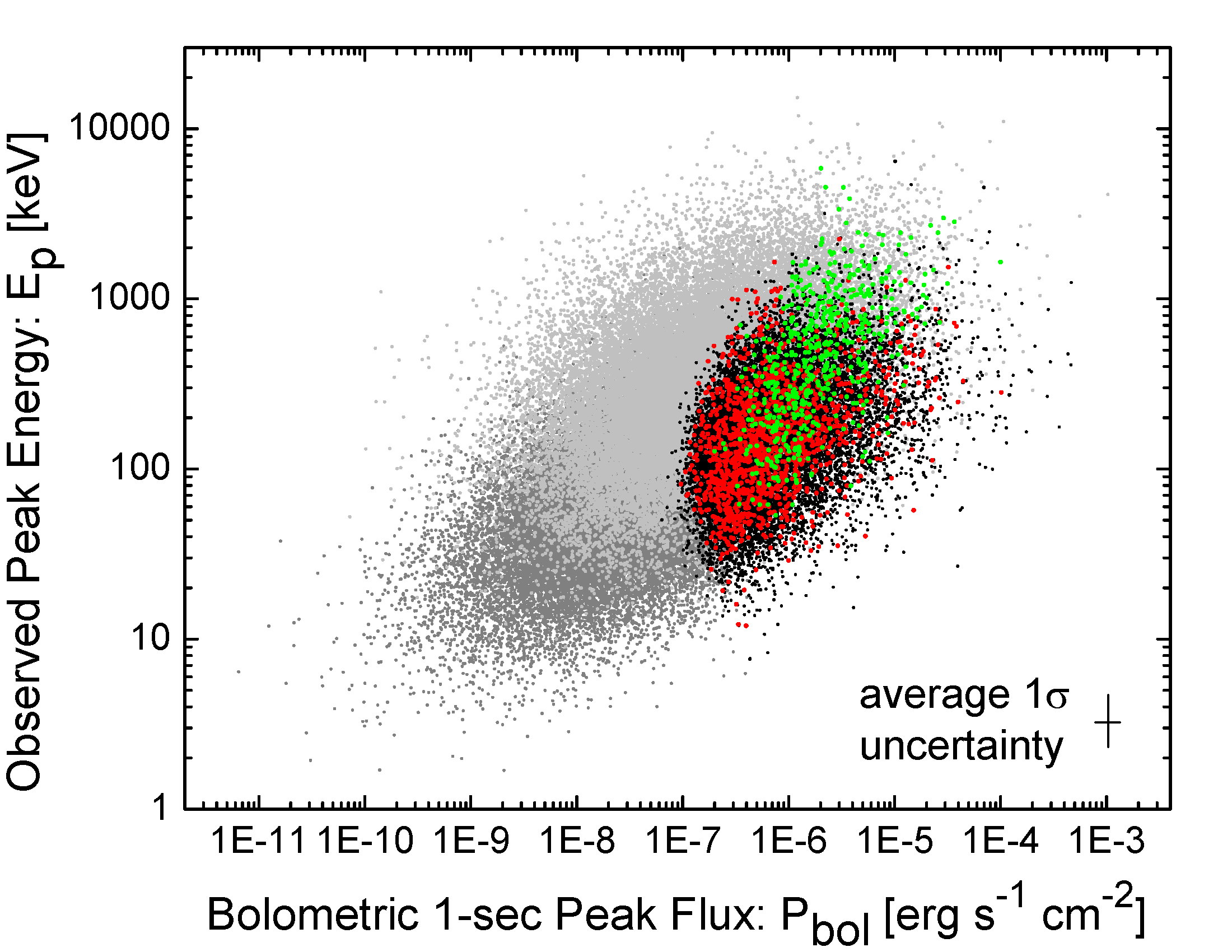} & \includegraphics[scale=0.309]{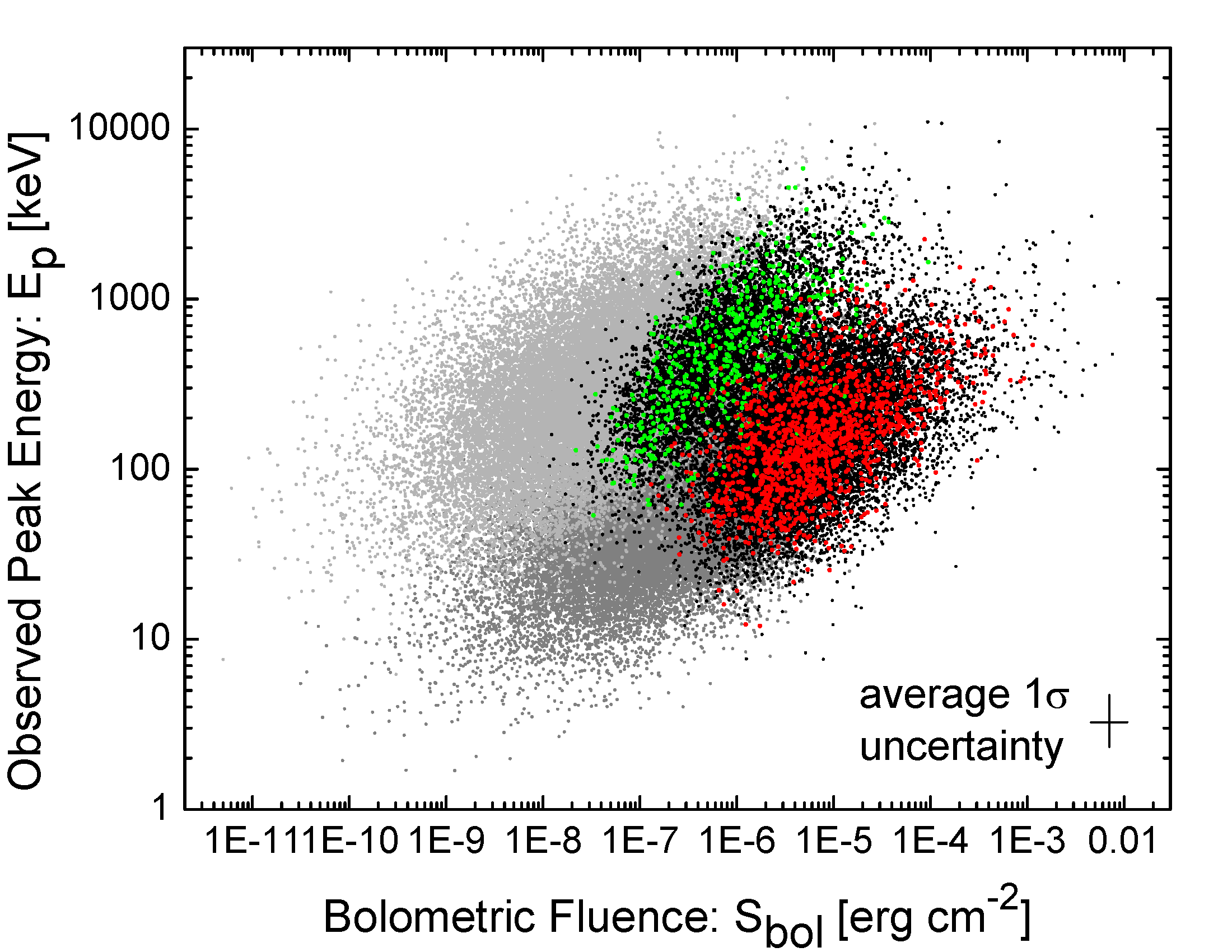} \\
        \end{tabular}
        \caption{\small Plots of BATSE $1366$ LGRBs and $600$ SGRBs prompt emission data superposed on the bivariate distribution predictions of the world models for the two classes of LGRBs and SGRBs. The statistical fitting is performed on the rest-frame parameters of GRBs by integrating over all possible redshifts while paying careful attention to BATSE detection efficiency. Extensive goodness-of-fit tests are provided in \citet{shahmoradi_multivariate_2013} and Shahmoradi \& Nemiroff (2013, in preparation). \label{fig:bivariate}}
    \end{figure}}

    \begin{itemize}

        \item
            The multivariate log-normal distribution provides excellent fit to population distribution of both BATSE Short \& Long GRB classes in the 4-dimensional prompt $\gamma$-ray emission parameter space of $\liso$, $\eiso$, $\epkz$ \& $\durz$ after taking into account the effects of detector thresholds and sample incompleteness \citep[c.f.,][also Shahmoradi \& Nemiroff, 2013, in preparation]{shahmoradi_multivariate_2013}.

        \item
            Despite potentially different progenitors of the two GRB classes, the preliminary results from the analysis of nearly 2000 BATSE catalog GRBs presented here indicate very similar interrelations and correlations among the four prompt variables ($\liso,~\eiso,~\epkz,~\durz$) of SGRBs and LGRBs. In particular, the correlation strength and significance of the Amati ($\eiso-\epkz$) and Yonetoku ($\liso-\epkz$) relations in the class of LGRBs are shown to be weaker than the previously reported values. The presented analysis, predicts a Pearson's correlation coefficient of $\rho\sim0.58\pm0.04$ and $\rho\sim0.46\pm0.07$ for the Amati and Yonetoku relations, respectively \citep[c.f.,][Sec. 3 \& Figure 7 therein for further discussion]{shahmoradi_multivariate_2013}.

        \item
            Based on a small sample of GRBs with known redshifts, the Ordinary Least Squares (OLS) regression slope of the Amati relation has been long known to be $\sim0.55$ \citep[e.g.,][]{amati_measuring_2008, ghirlanda_e_2008}. The LGRB world model here predicts the OLS regression slope of the Amati relation to be in the range $0.2-0.3$ after taking into account the effects of detector thresholds on the observed population of LGRBs (Figure \ref{fig:amati}). A similar analysis of the Fermi catalog of LGRBs recently done by \citet{heussaff_epeak-eiso_2013} reaches the same conclusion on the slope and dispersion of the Amati relation.

        \item
            Positive hardness-intensity correlations similar in strength to that of the Amati \& Yonetoku relations of LGRBs are also expected and predicted for the short class of GRBs, but with different normalization constants compared to LGRBs relations, shifting the SGRBs population to lower luminosities or output energies and harder spectral peak energies. The existence of positive correlations among the parameters $\liso$ \& $\eiso$ with $\epkz$ of both classes of GRBs can also be inferred from the population distribution of BATSE GRBs in the observer-frame parameter space (e.g., Figure \ref{fig:bivariate}, also Figure 8 in \citet{shahmoradi_hardness_2010}. This is mainly due to the fact that the majority of SGRBs \& LGRBs originate from the nearby universe with moderate redshifts $z\sim1-4$. Therefore, the convolution of the population distribution of the rest-frame parameters (i.e., $\liso$, $\eiso$, $\epkz$ \& $\durz$) with redshift distribution results in negligible variation in the shape of the joint observer-frame distribution of the same GRB parameters (i.e., $\pbol$, $\sbol$, $\epk$ \& $\dur$), leaving the relatively strong positive correlations among the four parameters almost intact in both observer and rest frame parameter spaces.

        \item
            The duration (e.g., $\durz$) of the prompt $\gamma$-ray emission in both classes of Short \& Long GRBs appears to be strongly correlated with the peak luminosity and the total isotropic emission of the bursts. The spectral peak energy of GRBs also correlates positively --but weakly-- with GRB duration \citep[c.f., Table 2 in][for quantitative measures of the associations for LGRBs parameters]{shahmoradi_multivariate_2013}. This positive trend is also evident in the observer-frame distribution of the parameters (e.g., Figure \ref{fig:bivariate}: {\it Center Right} plot).
    \end{itemize}

\Acknowledgements

I am very grateful to Robert Preece (at NASA$/$MSFC), Jon Hakkila (at College of Charleston), Istvan Horvath (at Bolyai Military University) \& Amy Lien (at NASA$/$GSFC) for helpful discussions.

\bibliographystyle{apj}
\bibliography{proc}

\end{document}